# Sunspot periodicity

Claudio Vita-Finzi
Department of Earth Sciences, Natural History Museum
London SW7 5BD, UK (*cvitafinzi@aol.com*)

6 December 2022

*The Schwabe (~11 yr) value for the annual sunspot number is sometimes uncritically applied to other measures of solar activity, direct and indirect, including the 10.7 cm radio flux, the inflow of galactic cosmic rays, solar flare frequency, terrestrial weather, and components of space climate, with the risk of a resulting loss of information. The ruling (Babcock) hypothesis and its derivatives link the sunspot cycle to dynamo processes mediated by differential solar rotation, but despite 60 years of observation and analysis the ~11 yr periodicity remains difficult to model; the possible contribution of planetary dynamics is undergoing a revival. The various solar sequences that genuinely display an ~11 yr cycle stand to benefit from an understanding of its periodicity that goes beyond statistical kinship. The outcome could ironically prompt the demotion of sunspots from their dominant historical role in favour of other possible indicators of solar cyclicity, such as the solar wind flux and its isotopic signatures, even if they are less accessible.*

> For what a man had rather were true he more readily believes
> Francis Bacon, *Novum Organon*, 1620

In 1843 the apothecary Heinrich Schwabe identified a periodicity in the sunspot record of about 10 years on the basis of sunspot groups for 1826-1850, a value endorsed by Alexander von Humboldt (1851) but soon revised to ~11 yr by Rudolf Wolf (1852a). Some kind of repetition had previously been recognised by Christian Horrebow in the number and size of the sunspots for 1761- 1776 (Fairbridsge et al. 2019), and for the period 1650-1800 by William Herschel (Fairbridge 1997a). For the last 4 centuries the annual total has ranged between 8 and 14 yr.

In due course an 11-year rhythm was found to ride at times on coarser periodicities derived from sunspot or $^{14}C$ data: the Hallstatt ~2400 yr cycle (Usoskin et al. 2016), the Suess-deVries ~200 yr cycle (Cini Castagnoli et al. 1998), and the Gleissberg ~ 88 yr cycle (Gleissberg 1958; Perystikh and Damon 2003). Several shorter cycles, including one measuring ~158 d (Oliver et al. 1998), have been identified for part of the sunspot or isotopic records.

Although it is not readily grasped without telescopic projection, the repetitive nature of sunspots attracted the interest of a wide range of astronomical, meteorological, medical and other researchers seeking to investigate solar influences on human health, economy and society (Herschel 1801, Jevons 1878, 1879; Hope-Simpson 1978; Hoyle and Wickramasinghe 1990; Juckett and Rosenberg 1993; Tapping et al. 2001; Towers 2017). Humboldt (1851) had seen no connection between sunspots and terrestrial climate; many others have continued to do so even if the causal chain between Sun and such matters as crop yield includes a number of what are decorously termed non-linear transition elements (Pustil'nik and Yom Din 2014 a, b). A disparate literature has resulted, with sunspots serving as the tokens of solar activity, just as natural selection was once so fully identified with its role as an evolutionary agency that it 'suffered neglect as an independent principle worthy of scientific study' (Fisher 1930).

This essay is motivated by the conviction that valid periodicities deserve to be identified because they would be useful in the analysis of the Sun's workings and by the same token that false cycles need to be rooted out.  I do not aim or claim to compete with the work of Hathaway (2015) or the countless workers at the sunspot coalface (the British Library catalogue alone lists 16,598 items under the heading of solar cycle.)

The paper is in three parts. The first part deals with the statistics and other numerical properties of the sunspot record. The second part considers some of the terrestrial and astronomical phenomena that are identified with an 11-yr cyclicity. The third part touches on a number of plausible mechanisms for solar periodicity including the role of planetary alignments. References are not necessarily the earliest mention of a phenomenon or process, the aim being to facilitate judicious access to a large and repetitive body of work.

**The ~11yr Sunspot Cycle**

It is no surprise, given the basis of its original identification, that statistical issues should loom large in the sunspot cycle literature.  The authors of the classic work on sunspots (Bray and Loughhead 1964, 226) were not impressed by the results achieved: 'the number of published investigations of a purely statistical nature would seem to exceed those having a more direct physical basis... [yet] the majority .... have led to no new insight into the physical nature of sunspots or the solar cycle'.

The present account is more sanguine and not only because there is an additional half century of material: besides advances in statistical methodology and computational devices we now benefit from three-dimensional views of sunspots and related features, the solar interior is being probed by helioseismology (Howe 2008), the associated magnetic and spectroscopic phenomena are well documented, and interpretation has been enriched by the fruits of solar system and stellar exploration, notably by the discovery of starspots and by numerical modelling of stellar interiors and galactic dynamics.

Even so the progress does not yet invalidate the comments made by Eddy (1977) in an early review that sunspots in themselves 'are not particularly fundamental in solar activity or necessarily related to the solar output', that 'we have little if any evidence that today's 11-yr cycle is an enduring feature of the sun', and that 'the available indirect evidence suggests that in the longer term, change and not regularity may rule solar behaviour'.

**Statistical scrutiny**

Sunspots, initially recorded simply as blots ('macchie') on the current solar hemisphere (Galilei 1613), were quantified by Wolf in 1848 using a formula ($N = k\,(10g + s)$) which combined the number of individual spots (*s*) and sunspot groups (*g*) into a single daily figure (*N*) (Wolf 1852a). This proved surprisingly reproducible despite the inclusion of a subjective value (*k*) for the observer and the place of observation. The formula was retained as the international sunspot number (SSN) or Zürich number until 2015, when the series, which includes historical data and runs from 1750 to the present, was subjected to a number of corrections necessitated by changes in equipment and in the weighting (*k*) adopted for the measurements (Clette et al. 2014).

The daily observations had also revealed that the solar body rotated monthly (Galilei 1613), that this rotation was latitudinally differential (Carrington 1863), and that it was accompanied by progressive equatorward shifts in the latitude of the spots over the Schwabe cycle (Carrington 1858, Spörer 1889). Once the magnetic nature of the spots had been revealed (Hale 1908) and the rhythm of the alternation in the polarity of sunspot pairs established (Hale and Nicholson 1938), the ~22 year Hale cycle came to be viewed by some workers as a related but distinct measure of solar cyclicity (Cliver 2014).

The statistical treatment to which sunspot records have been subjected ranges from the simplest to the most recondite. An example of the former is the averaging that led to 10 and then 11 years although even this procedure can be greatly refined, witness the value of 11.1212±6.0 favoured in one reference (Fairbridge 1997a). Solar cycle minima are found to exhibit the 80-120 yr Gleissberg cycle (Garcia and Mouradian 1998). Simple comparison brings out the variability that impedes even short-term prediction (Sabarinath and Anilkumar 2011). The comparative review of sunspot cycles by Wilson (1984) includes discussion of smoothed sunspot number and schematic sunspot cycles. Such surveys evidently benefit from extended sunspot records whether direct (Vaquero 2007) or derived from isotopic or other proxy sources (Usoskin 2013, Acero et al. 2018). The rates of growth and decay of the annual sunspot number have invited continuing scrutiny in order to establish their value for estimating future sunspot maxima and minima and their timing (Wilson and Hathaway 2006).

The length of a sunspot cycle, commonly defined by the appearance of the first spot of the new cycle, is complicated by the need to allow for an overlap of perhaps 1-2 yr between adjacent such cycles even though the lingering cycle operates at a different solar latitude (Kramynin and Mikhalina 2015). Alternative definitions include the interval between successive sunspot maxima and between the arrival at a specified latitude of average sunspot latitude drift in successive cycles (Miletsky and Ivanov 2014). Wavelet analysis (Ochadlick et al. 1993) may detect subtle variations in the length of the sunspot cycle and reveal changes in the prominence of putative 11-yr, 53-yr and 101-yr periods (Le and Wang 2003). Indeed, basing the analysis on minima or some other part of the sunspot number record may reveal the Gleissberg 80-120 yr period (Garcia and Mouradian 1998) and thus indicate one possible source of cycle length variability.

Compare those studies that were designed to discover whether variations in sunspot numbers are stochastic or chaotic. One such analysis (Morfill et al. 1991) dealt with the years 1878-1945 in units of 12 days, and it found that the record was best fitted by a 'deterministic chaos' model incorporating a memory on timescales of ~50 days. The finding has implications for magnetic field transport and perhaps also the mechanisms that produce and annihilate sunspots, but only if the sunspot generation is already known to be confined to the convective layer of the Sun. A later analysis assessed the relative importance of chaotic and stochastic behaviour in the course of cycles 10-23, that is during 1855-2008 (Greenkorn 2009). It found a change from stochastic to chaotic in the sunspot numbers between cycles 10-19 and 20-23 and boldly concluded that there was a corresponding increase in the scale of turbulence in the convective region.

Despite the complex interplay between stochastic, deterministic and chaotic elements in turbulence (e.g. Bershadskii 2011) such analyses have been applied in sunspot prediction, of increasing concern for such matters as the management of the orbits of artificial satellites by way of solar influences on atmospheric properties. Two major predictive strategies may be distinguished, one based on some kind of regression which allows extrapolation into the future, and a 'precursor' group of models which yields an estimate of the amplitude of the next cycle. The former hinges on cycle shape, as noted earlier, and the mathematical functions that convey it. Precursor methods generally employ sunspot numbers or geomagnetic indicators (Hathaway et al. 1999). Predicting the behaviour of a sunspot cycle is fairly reliable 3 years after the minimum in sunspot number (Hathaway et al. 1994).

Understanding the mechanism responsible for any periodicity (Spruit 2011) is evidently the key to more dependable forecasting. For instance, a combination of Laplace distribution functions and moving averages has served to model a generic sunspot cycle (Sabarinath and Anilkumar 2011), and what appeared to be a purely phenomenological characteristic of daily sunspot number - a Laplace or double exponential distribution of days on which the sunspot number changes - has been traced to changes in the subsurface magnetic fields which are otherwise out of sight (Noble and Wheatland 2013).The Babcock model, which hinges on the alternation between poloidal and toroidal configurations of the solar magnetic field, remains the ruling model even though it is sometimes dismissed as merely phenomenological and its validity may be questioned (e.g. Rabin et al. 1991).

Again, observation of sunspots as an evolutionary ensemble indicates the existence of a regular azimuthally directed subphotospheric magnetic field which changes its direction each 11-yr cycle (Ruzmaikin 2001). The crucial argument is that, whereas the field predicted by mean-field dynamo theory is too weak by itself to emerge at the solar surface, the turbulent character of solar convection means that the fields generated by the dynamo are intermittent and concentrated into ropes or sheets sufficiently strong to exceed the buoyancy threshold and emerge at the Sun's surface, with clustering of emerging loops to form sunspots. The classic model by Parker (1955) had also deduced many attributes of sunspots from the intersection of strands of magnetic flux from the solar toroidal field with the photosphere. More recently the strength of three sunspot cycles (21-23) was successfully predicted, using observations of the amplitude, penetration depth, equatorial return flow thickness and the position of the inversion layer of the meridional circulation, in order to calibrate a solar dynamo model (Choudhuri et al. 2007, 2017).

The Parker (1955) sunspot model mentioned above assumes a dynamo in which the requisite magnetic field originates at the tachocline and sunspots thus have their roots deep in the Sun's convection zone (Parker 1993). An alternative scheme views a dynamo that affects the convection zone as a whole and results in a reduction in the elasticity of the magnetic field lines leading among other things to 11-yr variations of the solar radius (Losada et al. 2016, Kleeorin et al. 1996). Now plotting the spectrum making up the sunspot frequency curve, which is consistent with the 22-yr periodicity, yields a large number of different frequencies rather than a simple sinusoidal curve (Dicke 1979b). This led to the suggestion that several practically independent processes were involved, a notion encapsulated in the eruption hypothesis in which each eruption takes about 11 years to die down and is guided by a random walk in phase (Kiepenhauer 1953). It was once argued that Babcock's theory of the solar cycle also yielded a random walk in phase, but various statistical tests for sunspot maxima in the late 18$^{th}$ century conflicted with that view and instead pointed to a precisely tuned internal oscillator acting as a clock within the Sun (Dicke 1978).

Grand solar minima and maxima are periods of several successive very low or very high Schwabe sunspot cycles, the best known being the 1645-1715 Maunder Minimum, first reported (as noted above) by E.W. Maunder, when sunspots almost vanished and there were few auroral sightings (Eddy 1976, 1983). Grand solar maxima can also be defined by a threshold of 600 MV (in 25-year averages) (Lockwood et al 2009). Solar-cycle modulation reportedly persists during extended periods of very low sunspot activity (Wang and Sheeley 2013), notably as $^{10}$Be concentration in the Dye 3 and NGRIP (Greenland) ice cores (Beer et al. 1998; Berggren et al. 2009).

**Individual sunspots**

Modelling the origin and life of individual sunspots (e.g. Alfvén 1943; Cowling 1945) has proved if anything a greater challenge than with pairs or gaggles of sunspots perhaps because the evidence lacks the context of a bipolar or migrating magnetic field. Nevertheless great advances in imaging resolution allow the fine structure of the sunspot (McIntosh 1981) to be monitored and its relation to photospheric granulation to be established. Sunspot pairs are an integral part of early attempts to systematise the polarity alternation that characterises the cycle (Akasofu 2014). They are central to the Parker (1955, 1975) model for sunspot formation in which concentrated flux tubes produced by the solar dynamo break through to the surface. There have been many elaborations of this suggestion (Rempel and Cheung 2014, Cheung and Isobe 2014); Caligari et al. (1995) concluded that the magnetic buyonacy to raise the tube required a field strength of ~$10^5$G. Novel approaches include the notion of negative effective magnetic pressure instability (Losada et al. 2016)

Changes in the intensity and magnetic field of individual spots are of course observed with ground-based instruments as well as from space (Albregtsen and Maltby 1981; Norton and Gilman 2004). The SDO/AIA and the Nobeyama radioheliograph revealed that the umbra of AR (Active Region) 11131 displayed 3-min oscillations in UV and EUV and in radio emission, with the higher frequency oscillations more pronounced near the centre of the umbra and the lower frequencies concentrated on the periphery (Reznikova et al. 2012). The evolution of line-of sight flows and horizontal proper motions in and around a follower sunspot, that is the trailing member of a sunspot pair, were documented over 5 d during which it decayed to a pore (Deng et al. 2008). Such measurements are useful guides to the underlying processes and thus conceivably those responsible for the 11 yr period, its variability, and its wider significance.

Pores, which are small sunspots lacking a penumbra (Sobotka 2003), do not necessarily evolve into fully fledged sunspots but spectral scanning may show an extended radial filamentary structure in the Ca II line with morphological and dynamical properties similar to those of the superpenumbra found in the chromosphere above large isolated sunspots (Sobotka et al. 2013). Flux magnetograms accordingly make it possible to identify pores, which may be difficult to detect on white light images and thus to monitor magnetic activity during pronounced solar minima, as in 2009 (Livingston and Penn 2009).

One component of sunspot behaviour that has resisted full understanding even though known for over a century, is Evershed flow, an outward flow of gas at up to 2 km/s from the sunspot's photospheric umbra which crosses the penumbra and sometimes extends beyond it and which was first identified spectroscopically (Evershed 1909; Rempel and Schlichenmaier 2011). An inward flow at higher altitudes -- the inverse Evershed flow (Beck and Choudhary 2019) -- consists of chromospheric material flowing into the sunspot along dark channels at about 40–50 km s$^{-1}$ (Georgakilas et al. 2003) with downflow over the umbra. Thanks partly to three-dimensional numerical simulation the fundamental mechanism is thought to be the radial component of thermal convection located mainly where the magnetic field strength is greatly reduced (Nordlund and Scharmer 2009) but even the basic observations - let alone their physical interpretation - are subject to dispute (Solanki 2003). However it is still surprising that peak photospheric flow velocities, unlike the magnetic field strength of sunspots, appear to be independent of the strength of the solar activity cycle (Penn and Livingston 2006, Hiremath 2010).

In addition to magnetograms and spectroscopic analysis of the chromosphere the study of individual sunspots has benefited from helioseismology, which bears not only on the tachocline and thus the genesis of the sunspot cycle (Bushby and Mason 2004) but also, in the guise of time–distance helioseismology, on conditions beneath the spot and the emergence of active regions (Kosovichev 1996; Thompson 2004).

**Sunspot distribution and the butterfly diagram**

For historical and related technical reasons the structure, dimensions and dynamics of sunspot groups are better known than those of their constituent spots. Consider for example the account by McIntosh (1981) built on 25 years of personal observation in white light. It considered sunspot groups and showed their intimate relation to photospheric granulation, mesogranulation and supergranulation. The axis of sunspot groups was found to begin at a high inclination to the equator which is gradually reduced by the addition of new spots. The velocity patterns revealed by H-alpha synoptic charts tevealed a sinusoidal zone of shear analogous to terrestrial jet streams. The study by Javaraiah (2011) focused on the growth rates of sunspot groups in 1874-2009 and found that the annually averaged daily mean growth rate was 70% greater than the corresponding decay rate, but, as a corrective to the blurring of averaging, a 11-year periodicity in both rates was found to apply mainly in the 0°-10° latitude zone of the southern hemisphere.

The magneto-hydrodynamic sunspot model advanced by Alfvén (1943, 1945; Walén 1944), which accounted for the generation of bipolar spots, hinged on two major tenets: that the darkness of the spot is a mechanical effect of its magnetic field and that this field is generated from the solar magnetic field by wave disturbances along the lines of force. It was dismissed by Cowling (1945) as internally inconsistent or inconsistent with observation.

The analysis by Schwabe (1843) was based on two annual features of the sunspot record, the number of spot-free days and the number of sunspot groups, and they yielded very similar sequences. Rudolf Wolf (1848) also took account of sunspot grouping in his formulation. The Sunspot Group Number count (Svalgaard and Schatten 2016) favours sunspot groups on the grounds that they would permit the record to be extended reliably back to times when telescopes were lacking or too defective for single spot recording (the backbone-method). Later work has sought to specify the properties of sunspot groups, including position in an 8-stage idealised development sequence (analysis of over 3000 sunspot groups by the Greenwich Observatory in the first half of the 20th century 50% have lifetimes of <2 d and 10% >11 d : http://www.sws.bom.gov.au/Educational/2/2/2 accessed 27 9 2019), doubtless primarily to increase their analytical value.

The group sunspot number ($R_G$) -- the number of groups that have been identified, normalised in order for it to agree with the Zurich sunspot number -- was introduced in 1998 (Hoyt and Schatten 1998) and has proved informative in revealing secular behaviour, as in the correlation between cycle number and cycle amplitude, and in extending the record by 4 early cycles (Hathaway et al. 2002). Of course 'normalisation' did not eliminate all discrepancies betwen the two sunspot series, and they have been reconciled by recalibrating the original international sunspot number SSN, the first time since its creation by Wolf, for the 400 yr period 1700-May 2015 (Clette et al. 2014). The revised series has a number of salient chacteristics. They include the absence of the 200-year deVries/Suess cyclicity that may be present in radionuclide datasets spanning millennia, raising the possibility that it is not a strictly solar effect such as a climatic influence on the mode of radionuclide deposition in polar ice (Ahluwalia and Ygbuhay 2016).

As we have seen, from the outset sunspot quantification, whether explicitly or not, dealt with the location on the solar surface as well as the number of sunspot assemblages (Li et al. 2003).. Perhaps the first attempt at formal sunspot mapping over a large area was the butterfly diagram pioneered by Edward and Annie Maunder (1904). The equatorward drift of sunspots that creates the butterfly effect operates largely at latitudes lower than 40°; above 40° there is a poleward migration of faculae. The plot vividly illustrated the repetitive pattern of sunspot shift and in passing the overlap between successive cycles and the asymmetry between the North and South hemispheres. With the extended solar cycle (ESC), whereby the wings of the activity butterfly extend nearly a decade back in time and to much higher solar latitudes (Srivastava et al. 2018), the periodicity measures 17-18 yr and two consecutive cycles with different levels of activity evolve simultaneously at different solar latitudes (Leroy and Noens 1983). In effect the ESC endorses the view that sunspot number and pattern reflect the emergence of magnetic flux rather than a more fundamental property of the Sun's activity.

The hemispheric asymmetry was noted both Spörer (1894) and Maunder (1904). The latter cautioned against any assumption to the contrary. The pattern is widely viewed as a useful test of competing dynamo models (Nepomnyaschikh et al. 2019), and investigations into the asymmetry focus on its amplitude or its phase (Hathaway 2015; Carbonell et al. 1993; Zolotova et al. 2009; Nair and Nayar 2008); but, despite great advances in computing power and algorithmic design, current MHD simulations still depart substantially from the observed solar cycle (Norton et al. 2014; Blanter et al. 2017).

Edward Maunder drew attention to the paucity of sunspots during 1645-1715, a period that is now called the Maunder Minimum (Eddy 1976); the few sunspots that prevailed during it were sufficient for Picard and La Hire to construct a butterfly diagram for 1615-1719 (Serre and Nesme-Ribes 1993, 1997). The fine structure of butterfly diagrams invites scrutiny both en masse or with reference to a limited wing flap. Kopecký and Kuklin (1969) focused on Cycles 12-19 with special reference to the double maximum displayed by 11-year sunspot data. Ternullo (2007) discussed Cycles 20-22 mainly in an attempt to decide whether the fine structure of the diagram is a real phenomenon or 'a meaningless consequence' of the turbulent nature of the eruption of magnetic flux, to find evidence of a periodic oscillation prograde and retrograde motion possibly linked to oscillation reported in the tachocline.

A further measure of sunspot distribution is the 'longitudinal asymmetry parameter' (Vernova et al. 2002) which is free from the stochastic, longitudinally even distribution of sunspot activity and thus emphasises the more systematic component of longitudinal asymmetry. It is argued that the stochastic fraction may account for the record height of cycle 19 and may be related to a relic magnetic field in the solar convection layer (Cowling 1946, Mursula et al. 2001). A relic dipole magnetic field in association with the Hale magnetic polarity cycle (Boyer & Levy 1984) is also invoked to explain the 22 year cyclicity in the Group Sunspot Number revealed for the last 400 yr with an amplitude of about 10% of the level indicated by present-day sunspot activity (Mursula et al. 2001, 2002).

Even though relic magnetic fields found elsewhere in the solar system (Vita-Finzi and Fortes 2013), other irregularities in the timing and duration of magnetic flux emergence, of which sunspots and active areas are of course symptoms, are rarely explained by appeals to relic effects. The 158-day periodicity in the occurrence of high-energy solar flares that has been identified by wavelet analysis of active regions (Oliver et al. 1998), for example, illustrates unexplained such complex timing. The major solar flares do not not observe the 11-yr or 22-yr cycle (Roy 1977).

**The ~11-yr Solar Cycle**

In the present account the term *solar cycle*, commonly viewed as synonymous with sunspot cycle (e.g. Eddy 1977), should be taken to encompass variations in the level of solar activity in general provided an ~11-yr periodicity which can plausibly be linked to the Sun has been demonstrated for the property in question. Some workers prefer the term 'solar activity cycle' for 'a near-periodic variation generally close to a period of 11 years' and take it to demonstrate 'the existence of a timing engine in the sun that appears to control or influence all aspects of solar phenomena and beyond in the solar system where the solar influence is felt' (Balogh et al. 2014). Others defer any such causal inference. Thus the association between sunspots and the declination of a magnetic needle reported by Wolf (1852b), like that between the Carrington coronal mass ejection of 1859 and a magnetometer disturbance at Kew Observatory (Cliver and Svalgaard 2004), was for some obscured by the hostility to any such link posed by Lord Kelvin, who pronounced that 'We may be forced to conclude that the supposed connection between magnetic storms and sunspots is unreal, and that the seeming agreement between the periods have been a mere coincidence' (Kelvin 1892; see also Stewart 1861).

The array of candidates for affiliation to the 11-yr solar cycle is diverse in content and plausibility although, as Lord Kelvin's contribution shows, authority (in this case *ab homine*) sometimes counts more than validity. Given that this review is rooted in the ~11-yr period first identified in the pattern of sunspots there would seem little point in hunting for putative solar cycle indicators on the Sun itself. Indeed, the fact that many of them are highly correlated with each other has been ascribed to their 'intrinsic link' with solar magnetism and consequently the 11-year cycle: they merely 'represent the many different observables modulated by the solar cycle' (Ermolli 2014). But this amounts to assuming that any variation on the photosphere or in the solar output (White 1977) is necessarily subject to the same (~11-yr) periodicity and mechanism.

**Solar indicators of the solar cycle**

Before considering instances of solar output as possible bearers of 11-yr cyclicity under the broad headings of matter, radiation and magnetic fields (White 1977) we need some discussion of a number of relevant properties of the Sun itself, namely its orbit and dimensions, as they could reflect and impinge on the solar cycle. The Milankovitch orbital (413,000, 125,000 and 95,000 yr) variables are too gradual for any 11-yr or 22-yr periodicity to be detected at present. However there are changes in solar spin configuration which are linked to progress along the solar cycle. For example the high-latitude migrating zonal flow pattern known as the torsional oscillation (Howard and LaBonte 1980) is associated with the rising phase of Cycle 24 (Howe et al. 2013).

Variations in solar dimensions have long been investigated. Changes in the Sun's diameter, measured telescopically from Earth or from space, have been correlated with sunspot number ever since the classic measurements by Picard and de la Hire in 1666-1719 (Ribes et al. 1987, Thuillier et al. 2005; Costa et al. 1999) and just as long challenged (Parkinson et al. 1980). Data for surface gravity (f-modes) obtained by SOHO and SDO instruments over two solar cycles (1996-2017) are thought to indicate changes in solar radius which are inversely related to sunspot number (Rozelot et al. 2018; see also Kuhn et al. 2012). Evidently secular constancy need not rule out short-term oscillations, and novel techniques, including the PICARD microsatellite programme and measurements at different wavelengths (Vita-Finzi 2013), may well prompt fresh disputes.

The solar interior has until recently been accessible only in terms of processes thought to account for energy production (Bethe 1939) and the gross structural divisions revealed by helioseismology (Ulrich 1970; Bahcall 1999), and thus eluded the hunt for periodicities. Not so the flux of solar neutrinos, assessed since 1964 (Davis 1964). The issue is obscured by a low instrumental count rate at the various neutrino detectors even though the flux is a very substantial $5.10^{10}$ s$^{-1}$ cm$^{-2}$. Data from both the Homestake and the GALLEX neutrino experiments point to modulation with a frequency of 11.85 yr$^{-1}$ The same frequency is prominent in power spectrum analyses of the ACRIM irradiance data for both the Homestake and GALLEX time intervals. Both could result from rotational modulation in the solar core, which rotates with a sidereal frequency of 12.85 yr$^{-1}$ (Sturrock 2009). Raychaudhuri (1984) had reported a statistically significant variation of solar neutrino flux data with the solar activity cycle, which he ascribed to the pulsating characters of the nuclear energy generation in the interior of the Sun, and 20 years of Homestake data (Lande et al. 1991) reportedly indicate that the average solar neutrino flux varies with the 11-yr solar activity cycle, as higher neutrino fluxes are observed during solar quiet periods and lower fluxes during solar active periods.

The detection of surface oscillations with a period of ~5 min (Leighton et al. 1962) in 1960 and their recognition as the impact of standing acoustic (p) waves trapped beneath the photosphere (Ulrich 1970) led to what came to be known as helioseismology. The mean frequency of low-degree p-modes in 1980-1985 was found to be higher at the time of solar maximum than near solar minimum (Woodard and Noyes 1985), possibly in response to changes in the strength of the solar magnetic activity near the solar surface. Moreover data for two solar cycles suggest that solar activity insofar as it is reflected in the modes varies from one cycle to the next (Broomhall et al. 2014).

Features identified on the photosphere by remote sensing, besides active areas and their outliers, include full-disk magnetograms taken at various ground observatories since the 1950s and from the SOHO and SDO satellites since 1996, and facular indices (1874-) at various observatories (Deng et al. 2013). Besides the magnetic fields manifested as sunspots and other components of active areas and the chromospheric network, the Sun boasts a global field which has been recorded as full-disk magnetograms from ground observatories since the 1950s and can be mapped by exploiting the Zeeman effect (1897) as in the Babcock magnetograph (Livingston 1999). Evidently observation of solar variation is not confined to the photosphere (or the core): possible chromospheric indices include full-disk Ca II K line and plage observations (Harvey 1992), daily full-disk H$_α$ observations (Makarov and Tlatov 2000) and a flare index (e.g. Kleczek 1952, Benz 2008). For the transition (1 GHz deemed better than the familiar F$_{10.7}$ index: Dudok de Wit 2014); and various coronal indices (e.g. Fe XIV emission line at 530.3 nm: Minarovjech et al. 2011). The magnetic flux emanating from the corona for 1964-1996, assessed from the aa geomagnetic activity index and interplanetary data (see below), rises and falls in each solar cycle lagging slightly behind sunspot number (Lockwood et al. 1999). Solar luminosity, defined as the Sun's total power outflow (Willson and Hudson 1991), when measured by the Active Cavity Radiometer Irradiance Monitor I (ACRIM I) during early1980-late 1989, varied with the 11-year solar cycle and closely followed statistical measures for the distribution of magnetic and photospheric features on the solar surface (Willson et al. 1986; Willson and Hudson 1988; Foukal and Lean 1990), pointing to the existence of a physical mechanism other than the thermal diffusion model that explains luminosity deficits due to sunspots. An exception was a strong irradiance excess during 1980 coinciding with the sunspot maximum of solar cycle 21 but not with any other indicators of solar activity.

Sometimes confused with luminosity, the Sun's total irradiance (TSI) -- the total electromagnetic energy spanning the entire spectrum per unit area above the Earth's atmosphere and normalized to one AU from the Sun -- has been measured over the years by radiometers on a variety of spacecraft (Li et al. 2010). The many periodicities (Lean 2018) embodied in the daily composite measure constructed by the Physikalisch-Meteorologisches-Observzatorium in Davos (PMOD) for 1978-2009 were found to include only two periods which were statistically significant following analysis by the complex Morlet wavelet transform method (Torrence and Compo 1998). They were 10.32 yr and ~32 d. The first was equated with the Schwabe cycle, the second with the 'quasi rotation period', thus called because its value appeared to differ from one cycle to the next. A revised value for the TSI of 1360.08±0.5 W m$^{-2}$ was obtained during the 2008 solar minimum using the Total Irradiance Monitor (TIM) on the Solar Radiation and Climate Experiment (SORCE) satellite (Kopp and Lean 2011), significantly lower than the value generally accepted in the 1990s (1365.4±1.3 W m$^{-2}$) mainly thanks to the exclusion of scattered light from the TIM (Kopp and Lawrence 2005). The TSI measurements show a difference of 0.12% between solar maximum and minimum. Variations in TSI lag behind sunspot activity by 29 days, in the view of Li et al. (2010), perhaps in response to the evolution of sunspots and bright features.

A definition of irradiance that complements TSI is spectrally resolved solar radiative flux (SSI:Yeo et al. 2014), again per unit area above the Earth's atmosphere and normalised to 1 AU. Both TSI and SSI have been monitored from space, at least in the UV wavelengths (120-400 nm), since 1978 (Kopp et al. 2012) but mainly because of instrumental difficulties there is more uncertainty over cyclical variability in SSI data and only the SIM instrument on the SORCE spacecraft spans the entire UV-IR spectrum and that only for 2003-2011. By 2014 the full period of record for ISS had not yielded any obvious solar cycle modulation (Yeo 2014). An exception is the replication by SSI of UV irradiance observations from the UARS and SORCE missions below 180 nm but as noted earlier the missions do not share identical viewpoints.

Until 1987 it was not certain whether variations in the UV component were subject to an 11-yr cycle, although there was evidence for their increase with solar activity during the solar cycle by an amount ranging from less than 1% at 300 nm to an order of magnitude at EUV coronal emission lines (Lean 1987). Continuous irradiance measurements in the EUV from 1991 by the Solar Stellar Irradiance Comparison Experiment (SOLSTICE) and the Solar Ultraviolet Spectral Irradiance Monitor (SUSIM) instruments on the Upper Atmosphere Research Satellite (UARS) satellite soon allowed the 11-year variability of the Sun's radiation at wavelengths 120-400 nm to be estimated (Lean et al. 1992; Rottman 1999). Further progress came with the Solar EUV Experiment's EUV Grating Spectrograph (EGS) onboard the Thermosphere Ionosphere Mesosphere Energetics Dynamics (TIMED) spacecraft which has provided low-resolution spectral data since 2002 (Woods and Eparvier 2006; Dudok de Wit et al. 2008).

By now it will be clear that the search for 11-yr cyclicity - or its rejection - encompasses a wide range of wavelengths from radio to gamma rays, and that reconciling divergences between measurements and both proxy and semi-empirical models is hampered by persistent uncertainties (Yeo et al. 2014). Yet it is sometimes facilitated by appeals to time lags for which there is no robust justification. Take the reported phase asynchrony between sunspot numbers and the coronal index (CI), which refers to the total energy emitted by the E-corona at 530.3 nm (Fe XIV) (Mavromichalaki et al. 2005). The correlation coefficient between the two series varies between cycles, but on average the SNs begin a month earlier in the period January 1939-December 2008 (Cycles 22 and 23: Deng et al. 2012), an example of phase asynchrony ascribed to their different origins (Deng et al. 2012) even though for the same period a periodicity of ~ 11.45 yr was found by Fast Fourier Transform and Morlet Wavelet Transform for CI, GCRs and also a solar flare index (SFI) (Singh et al. 2018).

Synthetic versions of particular wavelengths may in fact reveal or rule out disputed explanations. The $E_{10.7}$ proxy for EUV (Tobiska 2001) represents the integrated solar EUV energy flux for 1.8-105.0 nm at the top of the atmosphere in units of 10.7 cm radio flux It has the advantage of extending back to February 1947 and therefore spanning potentially more than one solar cycle. An ~11-yr period is indeed clearly represented for 1992-2002 (Vita-Finzi 2009) but the data also embody a sequence of ~27-day oscillations which, like the neutrino record from the Homestake detector, hints at an origin in rotation of the solar radiative zone (Sturrock et al. 1997) rather than the passage of active regions across the solar surface (Donnelly 1987).

The solar output is by no means confined to continuous fluxes and includes ephemeral, sometimes violent, events such as solar storms and the emission of plasma by coronal mass ejections (CMEs). Edward Sabine (1852) noted that the 11-yr sunspot period corresponded with a similar period in magnetic storms and other features of terrestrial magnetism, and it has since long been reported that solar flares and CMEs as well as storms are most frequent at or near solar maximum (Webb and Howard 1994; Ramesh 2010; Le et al. 2013; Telloni et al. 2014, 2016; Li et al. 2018).

**The heliosphere**

The solar wind, which emanates from the corona, also reflects the solar cycle (Neugebauer 1975, Richardson and Kasper 2008,Veselovsky et al. 2012) and may in turn influence its dynamics (Israelevich et al. 2000). It defines the heliosphere (Gazis 1996; Tokumaru 2013). Using data obtained by the Wind spacecraft, Kasper et al. (2012; see also Aellig et al. 2001) found that over Cycle 23 between 1994 and 2010 the abundance of helium relative to hydrogen in the slow strain of solar wind was strongly correlated with sunspot number (Ogilvie and Hirshberg 1974, McIntosh et al. 2011). This association gains practical importance in view of the suspected link between helium content in the solar wind speed and the incidence of coronal mass ejections (Richardson et al. 2003); how far it depends on changing sources on the Sun (Poletto 2013) is uncertain.

What is well attested is that the solar wind acts as shield against the flux of galactic cosmic rays (GCR) from interstellar space (van Allen 2000; Cliver et al. 2013; Bazilevskaya et al. 2014). Instrumental monitoring of the GCR record from ionization chambers and neutron monitors has revealed a clear ~11-yr periodicíty for anti-correlation with solar activity, with sunspot number as proxy (Forbush 1958; Neher 1967; Krainev and Kalinin 2013; Ross and Chaplin 2019). For earlier times the cosmogenic isotope $^{10}$Be in ice cores also yields 11-year variations albeit of greater amplitude because $^{10}$Be is more sensitive to lower energy cosmic rays than are neutron monitors (Beer et al. 1990, McCracken 2001, 2004), although the $^{10}$Be signal in ice cores may be distorted by complex ice flow (Berggren et al. 2009). The atmosphere acts as a low-pass filter so that long-term variations are emphasised at the expense of short-term events. In addition solar flares can increase the $^{14}$C in tree rings by about 1% (Kocharov 1992). Various studies have accordingly found no persuasive 11-yr signal (Suess 1965); a rare exception is an analysis of wines produced in 1909-1952, which showed $^{14}$C variations with an amplitude of 4.3±1.1 ‰ (Burchuladze et al. 1980).

The Ulysses solar mission 17 years into its 22-yr Hale cycle had observed that the magnetic field of the heliosphere was organised into two regions of opposite polarity separated by a single current sheet (Balogh et al. 2001; Fisk and Zhao 2009; Zhao and Fisk 2011). This became tilted with regard to the solar equator as the cycle progressed and eventually 'turned over'. The asymmetry in solar wind speed distribution also alternates with the Hale cycle (Mursula and Zieger 2001).

**The terrestrial record**

The major source of information on the Sun's magnetic history is thus terrestrial. The longest record (1868-) is held by the *aa* geomagnetic index (Mayaud 1972; Thompson 1993; Coffey and Erwin 2001; Lockwood et al. 2018), a measure derived from the K index (a measure of disturbances in the horizontal component of the Earth's magnetic field) for two roughly antipodal observatories every three hours. Some of the cyclicity it displays may derive from orbital intersection with high-speed solar wind streams (Zerbo et al. 2013) rather than from an intrinsic feature of the magnetic signal. However, over the period 1844-1994, i.e. over six Hale cycles, the amplitudes of the 22-year sunspot and geomagnetic activity cycles were found to be highly correlated (Cliver et al. 1996). Other terrestrial indicators with apparent 11-yr cyclicity include the Earth's spin rate (Currie 1980, 1981) which, to judge from sunspot and Doppler data, and latterly the findings of helioseismology, varies with the solar cycle by about 1% especially at the solar equator (Howard 1984; Tuominen and Virtanen 1987; Hathaway and Wilson 1990; Bhatnagar and Livingston 2005).

The most vociferous and persistent terrestrial echo of the 11-yr cycle is of course climate (Huntington 1945; Soon et al. 2014) with its many hydrological, biological and economic ramifications. Needham (1959, 435) observed that, although a connection between sunspot activity and the conditions of the ionosphere was fully accepted, its effects on the meteorological situation was uncertain, quoting Ionides and Ionides (1939) to the effect that, for once, Chinese astrological conclusions were perhaps right in maintaining a connection between celestial and terrestrial phenomena. 'Though not at present generally admitted it is not impossible that the sun-spot period may be connected, through meteorological effects, with events of social importance, such as good and bad harvests', a view maintained by Arakawa (1953) about Japanese rice harvest famines since 1750.

Even when the solar factor appears inadequate to account for its alleged impact (e.g. Kelly and Wigley 1990) and in the face of repeated warnings against spurious correlation (Pittock 1978; Baldwin and Dunkerton 1989), the hunt for an 11-yr signal in the atmosphere remains active and is sometimes rewarded if only after judicious targeting. Meteorological data from which volcanic and El Niño signals have been purged reveal global scale tropospheric changes consistent with solar forcing (Gleisner and Thejll 2003); total column ozone may be implicated not by direct radiative interaction but by the solar influence on the poleward transport of the gas (Labitzke and van Loon 1997); and the strong correlation of sunspot cycle length with northern hemisphere temperature anomaly (Friis-Christensen and Lassen 1991) shows how matching irregular sequences, perhaps helped by the introduction of a neglected factor such as low cloud cover (Svensmark and Friis-Christensen 1997), can prove a valuable investigative device when the causal links remain uncertain even after experimental scrutiny (Carslaw et al. 2002; Zarrouk and Bennaceur 2010; Kirkby et al. 2011).

Computer modelling of general circulation models is also a route to the testing of the 11-yr periodicity (Meehl et al. 2009). One can settle on a limited number of controlling variables, such as solar irradiance and stratospheric ozone (Haigh 1996), and the outcome may in turn lead to changes in the large-scale tropospheric circulation which are amenable to testing, such as changes in storm-track location related to the 11-yr cycle (Vita-Finzi 2018). Even more ambitiously, evidence of the 11-yr rhythm has been identified in climatic 'measures of variability' such as the North Atlantic Oscillation (NAO), the southern annular mode (SAM) and the El Niño–Southern Oscillation (ENSO) which encompass substantial parts of the globe (Anderson 1990; Landscheidt 2000; Kuroda and Kodera 2005; Barriopedro et al. 2008; Li et al. 2011; Hassan et al. 2016), although the methods employed and the underlying assumptions have their critics (Gray et al. 2010).

To correlate individual weather components with the solar cycle would appear more straightforward (Hiremath 2006), but this does not guarantee a convincing interpretation. For instance, a drought cycle with a 22.36 yr periodicity spanning 1000 years has been derived for the western USA from the D/H isotope ratio in the cellulose of Bristlecone pine tree rings (Epstein and Yapp 1976); but, rather than a climatic effect, Dicke (1979a) favoured as explanation variations in solar luminosity driven by a deeply buried magnetic field. With the less specific air temperature, classical spectrum analysis had been supplanted by maximum entropy spectrum analysis (MESA); when applied to the entire globe it revealed the 11-yr signal only in North America east of the Rockies and North of 35° N and it indicated a solar cycle period with a mean amplitude of 0.3°C and lagging behind sunspot numbers by $5.7 \pm 0.7$ yr (Currie 1981). Tide gauge data is evidently further removed from solar control, which is perhaps why a 10.6 yr cycle with an amplitude of 10-15 mm was found in mean sea level data for Europe, in anti-phase to the sunspot cycle in high latitudes but approximately in phase at mid-latitudes (Woodworth 1985).

Similarly, although surface hydrology is governed by geological and vegetational as well as climatic factors the simplicity of river discharge totals invites correlation with one or other index of solar activity. Fourier analysis of Nile flood levels for AD 622–1470 yielded a period of 18.4±0.4 yr, which was attributed to the lunar tide of 18.6 yr (Hameed 1984); yet Empirical Mode Decomposition showed that the 11-yr cycle is present in the river's high-water level variations which reflect solar influence on the Nile sources in eastern Equatorial Africa by way of the atmospheric circulation over the Atlantic and Indian Oceans (Ruzmaikin et al. 2006). The role of galactic cosmic ray flux is another blur in the causal web between solar cyclicity and weather (Marsh et al. 2000) which is gaining experimental as well as observational support (Yamaguchi et al. 2010; Shaviv 2005; Kirkby 2011).

The agricultural impact of weather and climate may seem even less likely to yield robust conclusions thanks to the confusions introduced by the human factor. Yet it yields mappable and archival data often well calibrated over time and, as the survival of many farming economies (and the taxes they yield) shows, systems that may last decades and centuries. It was here that some of the earliest conjectures about solar fluctuations arose (Herschel 1801; Jevons 1882; Lockyer et al. 1896) and that disputes over their validity continue to flourish (Love 2013). Indeed, the term sunspot is employed for an extrinsic random variable in economics (Cass and Shell 1983). In any case the association between solar periodicity and crop yields has been given fresh credence by the analysis of the connection between agricultural prices and space weather by way of a causal chain which includes what are delicately called several non-linear transition elements (Pustil'nik and Yom Din 2004a and b).

**Implications**

Much has been discovered about the history of the ~11-yr periodicity from written records and through the medium of cosmogenic isotopes (Baliunas and Jastrow 1990; Anderson 1991; Cini Castagnoli and Provenzale 1997; Usoskin 2013; McCracken et al. 2013; Arlt and Weiss 2014). Yet it seems premature to assume that the 11-yr solar activity cycle is 'the observational manifest of the solar dynamo' (Yeo et al. 2014) even if a less restrictive claim can more confidently be made that 'the solar field is maintained by a hydromagnetic dynamo' (Tobias 2002). As it happens it was the irregularity of the cycle period that prompted the notion of a 'chronometer' deep in the Sun (Dicke 1979b): irregularity is not the attribute that leaps to mind when considering a decent chronometer or, indeed, a working dynamo.

The dynamo model as generally understood (Charbonneau 2010, 2014) hinges on differential rotation and meridional flow (Weiss and Thompson 2009); the one enduring model for the dynamo-cycle is the empirical one by Babcock (1961), which invokes flux loops to account for the alternating cycle, a stratagem rendered implausible by a solar surface impervious to magnetic flux (Parker 1984). Like the related semi-empirical model devised by Leighton (1969; Charbonneau 2007; Cameron and Schüssler 2017), the solar dynamo theories of Babcock and Parker have both been dismissed because the requisite diffusion coefficient was lacking (Piddington 1972) or because an axisymmetric magnetic field cannot be sustained by dynamo action (Cowling 1934, 1957).

Whereas in creating those models and their derivatives it was generally assumed that the dynamo action was distributed throughout the Sun's convection zone or located near the photosphere, there is currently some support for it to be concentrated at or near the tachocline, the interface between the convective and radiative zones . No convincing evidence for a mechanism to drive an 11-yr cycle has yet come to light thereabouts (Tobias and Weiss 2007), but in seeking to explain the periodicity and its irregularity it is helpful to partition the sequence among the documented physical processes operating in the Sun. In one such fragmentary scheme (Mullen 2010) the successive episodes total 9-12 yr.

**Mechanisms: external**

A case has long been made for a contribution by planetary alignments to the solar cycle (Schuster 1911). The association proposed in China in the 3$^{rd}$-4$^{th}$ century BC between the Jupiter cycle and agricultural production may in fact have been occasioned by the 11-yr sunspot period (Needham 1959). Wolf (1859), among others, remarked on the closeness of the 11-yr sunspot cycle to Jupiter's sidereal period and the combined effect of Venus, the Earth, Jupiter and Saturn (Scafetta and Bianchini 2022); a link between sunspots and Jupiter was explicitly discussed by Carrington (1863) and between Mercury and sunspots a century later by Bigg (1967). The planetary hypothesis lost favour once the magnetic nature of sunspots had been revealed by Hale (1908; Charbonneau 2013) but never completely. Besides recalculation made possible by astronomical and computing advances there have been innovations in the analysis as a whole, notably in response to the revelations of helioseismology (Dingle et al. 1973; Hughes et al. 2007).

How the control is accomplished has not always been specified, but gravitational and derived tidal factors are the natural corollary of a Newtonian solar system, with a centre of mass, or barycentre, which is seen as the crucial component of any such system (Jose 1965; Hung 2007; Okhlopkov 2012; Fairbridge and Shirley 1987; Scafetta and Willson 2013), and the amplitude of the 11-yr cycle alternates between phases of high and low activity (the latter including the Maunder and other minima) according to position along an orbit determined by the solar system as a whole (Charvátová 1988).

A more indirect route which accepts that the planets cannot directly control solar activity but perturb the operation of the solar dynamo posits that they exert a time-dependent torque on the solar core (Grandpierre 1996) or on a non-spherical tachocline at the base of the convective zone (Callebaut et al. 2012). The resulting periodicities were found to tally with long-term cycles in proxies of solar activity over 9400 years derived from cosmogenic isotopes (Abreu et al. 2012). However, the statistical basis for the thesis has been found wanting in two ways: the purported peaks in the torque spectrum are artefacts produced by aliasing, while any coherence between planetary torque and 'heliospheric modulation potential' is in fact insignificnt (Polouianov and Usoskin 2014). More recently, the tidal-forcing option has been revived, with Venus, Earth and Jupiter joining forces to influence the solar magnetic field (Takahashi 1968; Stefani et al. 2019, 2016). A Venus-Earth-Jupiter alignment occurs every 11.07 yr, an effect hitherto dismissed as inadequate but now reinforced by reference to the Tayler instability, a component of the Tayler-Spruit dynamo model for the generation of stellar magnetic fields (Rogers 2011; Stefani et al. 2018). And electrodynamic rather than gravitational effects have been invoked to link planetary configuration with solar activity (Nikulin 2009; Scafetta and Bianchini 2022).

The identification of sunspot cycles on other stars (Noyes 1991) has proved beneficial to planetary models for solar periodicity not only by encouraging the notion of solar system dynamic interaction but also by making available data on age, mass, rotation and other characteristics of Sun-like stars to complement our parochial database. The inference that stellar light variability represents spots was first advanced in 1972 and circumstantial support for it includes rotation and periodic shifts in their latitude. Moreover the putative starspots hint at active magnetic activity. In 2009 it was still possible to ask 'whether starspots are just blown-up sunspot analogs' (Strassmeier 2009). Before long the data amassed by the Kepler space telescope mission were complex enough to raise such issues as differential surface rotation and spot emergence and decay (Basri 2018). From our present perspective what renders them significant is indeed their analogy with solar sunspots and in particular their display of periodicities in latitude and magnitude. The COROT light curve for CoRot-Exo-2, for example, has a period of 4.522 d indicative of two spots at different latitudes (Strassmeier 2009 );17 yr of photometry of G8 giant HD 08472 shows a brightness cycle of 6.2 yr, a possible stellar analogy of our 11-yr cycle (Özdarcan et al. 2010).

Besides illustrating the possible links between cycle length and the nature of the host star, such stellar analogies help to guide extrapolation of the limited observational data into the future and the distant past. Indeed, the predictive limitations of the Babcock-Leighton model (Bushby and Tobias 2007; Charbonneau 2007b; Cameron and Schüssler 2017) and the prevailing algorithms (Pesnell 2015), and the growing practical demands and risks of space exploration (Cucinotta and Durante 2006; Goelzer et al. 2013; Schwadron et al. 2018), doubtless encourage renewed assessment of planetary and other extrasolar influences (e.g. Kotov et al. 2012) on the cycle period.

It is not solely a matter of timing. For example, the 11-yr modulation varies by a factor of about 5 for protons of ~200 MeV and only a few percent for energetic ~10 GeV cosmic ray muons (Foukal 2013); long-term solar brightness changes may be estimated from Sun-like stars (Lockwood et al. 1992). Invoking extrasolar influences on the solar cycle thus opens the door to analysis of the solar dynamo in terms of stellar generalities. It may also lead to the dethronement of sunspots and their replacement by other indicators of solar cyclicity, such as the solar wind flux, which, though less accessible, benefit from records spanning many millennia and from an explicit link to the dynamics of the solar interior.

 **Summary and conclusions**

Since its recognition in 1843 by Heinrich Schwabe the ~11-yr sunspot cycle, often termed the solar cycle, has been identified in periodicities displayed by various components of the Sun's radiation and other parts of its output and by related phenomena in space and on Earth. Although in some instances the ~11-yr solar equivalence is endorsed by calculation or modelling, many accounts present minimal analysis and invite the charge of wishful thinking. As a consequence attempts to explain the sunspot cycle itself are dominated by models which invoke some aspect of the dynamo that is thought to operate in some part of the Sun's convection zone. To be sure, the few current alternatives, including those that rely on planetary alignments, are unconvincing though doubtless in part through the taint of astrology, and there remains the validity of equating with the solar cycle – itself with a duration of somewhere between 8 and 14 yr – a range of climatic, geological and medical episodes many of which lack independent chronologies. In the enunciation of the laws of astronomy, as well as physics and mechanics, commented Henri Poincaré, we choose rules whereby 'The simultaneity of two events, or...the equality of two durations...are only the fruit of an unconscious opportunism' (Gould 2001). This cynical assessment can be invalidated by applying chronometric techniques that were not available in 1905 when its author made his claim.

## Acknowledgments
I thank two anonymous referees for their forthright comments on the MS.